\documentclass[%
 reprint,
superscriptaddress,
showpacs,preprintnumbers,
nofootinbib,
 amsmath,amssymb, 
 aps,
 prd,
 longbibliography,
]{revtex4-1}

\usepackage{cancel}
\usepackage{accents}
\usepackage{mciteplus,slashed}
\usepackage{amssymb,cancel,amsmath}
\usepackage{dcolumn}
\usepackage{bm}
\usepackage[caption=false]{subfig}
\usepackage{appendix}
\usepackage{physics}
\usepackage{booktabs}
\usepackage{feynmp-auto}
\usepackage[T1]{fontenc}	
\usepackage{csvsimple}
\usepackage[hidelinks]{hyperref}
\usepackage[section]{placeins}
\usepackage[capitalise]{cleveref}
\usepackage{booktabs}
\usepackage{graphicx}
\usepackage{mathrsfs}
\graphicspath{{Figures/}}

\AtBeginDocument{
\heavyrulewidth=.08em
\lightrulewidth=.05em
\cmidrulewidth=.03em
\belowrulesep=.65ex
\belowbottomsep=0pt
\aboverulesep=.4ex
\abovetopsep=0pt
\cmidrulesep=\doublerulesep
\cmidrulekern=.5em
\defaultaddspace=.5em
}
\usepackage[dvipsnames]{xcolor}
\usepackage[normalem]{ulem}
\usepackage{fontawesome} 

\begin{document}

\title{Coulomb Corrections for Coherent Neutrino Nucleus Scattering}

\author{Ryan Plestid}

\affiliation{Walter
 Burke Institute for Theoretical Physics, California Institute of Technology, Pasadena, CA 91125}

\preprint{CALT-TH-2023-008}

\begin{abstract}
 In this work we consider sub-leading $O(Z^2\alpha^3)$ corrections to coherent elastic neutrino nucleus scattering (CEvNS). These corrections are not negligible by power counting since nuclei with large coherent cross sections have sizeable nuclear charges e.g.\ $Z\alpha \sim 0.4$. We find that the corrections are much smaller than naive power counting in $Z\alpha$ would suggest and that Coulomb corrections do not substantially alter predictions for CEvNS in the Standard Model. We comment on similarities to older literature on mesonic and muonic atoms. 
\end{abstract}

 \maketitle

\section{Introduction} 

Coherent elastic neutrino nucleus scattering \cite{Stodolsky:1966zz,Freedman:1973yd,Kopeliovich:1974mv} (CEvNS) offers a clean probe of the weak neutral current. It is a leading tool for constraining neutrinophillic models of new physics \cite{deNiverville:2015mwa,Papoulias:2019xaw,Billard:2018jnl,Altmannshofer:2018xyo,Abdullah:2018ykz,Blanco:2019vyp,Farzan:2018gtr,Miranda:2020syh}, and it furthermore supplies valuable information on nuclear structure \cite{Amanik:2007ce,Huang:2019ene,Coloma:2020nhf,VanDessel:2020epd,Abdullah:2022zue}. CEvNS was recently measured for the first time \cite{Akimov:2017ade} and there now exist dozens of experimental collaborations that will attempt to measure a CEvNS signal in the near future \cite{Abdullah:2022zue}. The COHERENT collaboration, in particular, has aspirations of sub-percent level precision (see e.g.\ Fig.\ 19 of \cite{Akimov:2022oyb}). Similar projections of percent level measurements have been proposed for reactor neutrinos with low-threshold detectors \cite{Fernandez-Moroni:2020yyl}.

\begin{figure} 
    \vspace{12pt}
    \includegraphics[width=0.6\linewidth]{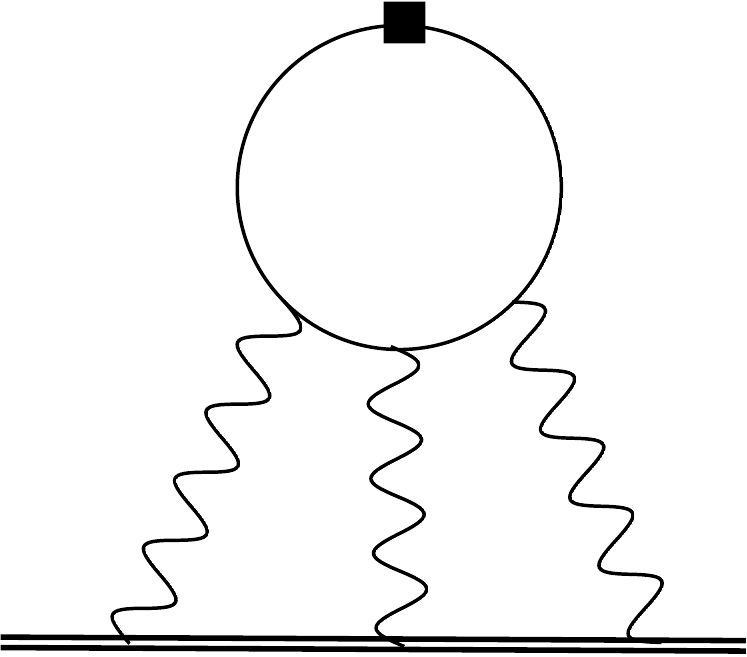} 
    \caption{Graph that contributes to CEvNS considered in this work. The square denotes a four-Fermi contact operator (neutrinos not drawn) and the double line the nucleus. In the infinite mass limit the nucleus behaves as a static potential of charge $Z$. The loop is any dynamical charged particle in the effective theory e.g.\ muons, electrons, or pions.  In practice only electrons are relevant phenomenologically. \label{diagram} } 
\end{figure}

Motivated by recent interest in precision measurements of the CEvNS cross section, the authors of \cite{Tomalak:2020zfh} performed a study of radiative corrections to CEvNS at $O(\alpha)$ (see also earlier work \cite{Sehgal:1985iu,Botella:1986wy} along these lines). The calculation was performed in a low energy effective field theory (EFT) in which any large ``RG-logs'' had already been resummed into Wilson coefficients \cite{Hill:2019xqk}. Accounting for hadronic, nuclear, and perturbative uncertainties, the authors of \cite{Tomalak:2020zfh} concluded that CEvNS represents a bonafide precision observable within the SM. In this work I consider Coulomb corrections to the charged lepton loop that mediates photon exchange with the nucleus; the diagram of interest is depicted in \cref{diagram}. This constitutes the only relevant radiative corrections that was not considered previously; other corrections exist but are suppressed by higher powers of $\alpha$ with no coherent enhancements. Formally, \cref{diagram} represents a relative $O(Z^2\alpha^2)$ correction to the results computed in \cite{Tomalak:2020zfh} (which are themselves $O(Z\alpha)$). For heavy nuclei this relative correction could be sizeable. As we will now discuss, however, the correction is actually much smaller than naive power counting would suggest and can be safely neglected for practical purposes. 

The rest of the paper is organized as follows: \cref{Coulomb-Corrections} discusses how Coulomb corrections enter in CEvNS and reviews existing results on vacuum polarization in an electrostatic field. Next in \cref{Results},  I compute Coulomb corrections from lepton loops for CEvNS in the point-like limit making use of existing work from the muonic atom literature; the effect is found to be small. In \cref{Discussion} I discuss finite size corrections and hadronic vacuum polarization and argue these effects, while interesting theoretically, are not relevant for a percent-level determination of the cross section. Finally in \cref{Conclusions} I summarize, conclude, and comment on the position of CEvNS as a precision SM observable. 

\section{Coulomb corrections \label{Coulomb-Corrections}} 
CEvNS is a particularly simple system from the perspective of radiative corrections due to the absence of charge particles on the external legs. Nuclei carry macroscopic charge, however their coupling to real photons is mass-suppressed\footnote{This is most easily seen by using a heavy particle Lagrangian and working in Coulomb gauge.} and their only influence at leading order in $1/M$ is via the exchange of Coulomb photons. Short-distance radiative corrections can be incorporated via standard EFT techniques. The only residual contribution that must then be considered involve loops of light particles. 

In what follows we discuss radiative corrections at $O(Z^n \alpha^{n+1})$. This restriction allows us to express the amplitude as 
\begin{equation}
    \mathcal{M} = \mathcal{M}_0 + \frac{\alpha}{\pi}  \sum_n (Z\alpha)^{2n}\mathcal{M}_1^{(n)}~,
\end{equation}
Only even powers of $Z\alpha$ appear by Furry's theorem \cite{Furry:1937zz}. The tree-level amplitude $\mathcal{M}_0$ is proportional to the weak nuclear form factor, $F_W(Q^2)$,  whereas all $O(\alpha)$ corrections  in  $\mathcal{M}_1^{(n)}$ depend only on the charge form factor, $F_{\rm ch}(Q^2)$. 

At one-loop the amplitude is modified in a simple way that depends only on the QED vacuum polarization \cite{Tomalak:2020zfh}
\begin{equation}
    \mathcal{M}_1^{(0)} = \frac{\alpha}{\pi} F_{\rm ch}(Q^2) (\delta^{\nu_\ell} + \delta^{\rm QCD}) ~, 
\end{equation} 
where $\delta^{\nu_\ell}$ is the sum of vacuum polarizations $\Pi(Q^2,m_\ell)$, weighted by flavor dependent Wilson coefficients. The correction $\delta^{\rm QCD}$ is likewise related to hadronic vacuum polarization, and a charge isospin correlator \cite{Jegerlehner:1985gq,Tomalak:2019ibg,Tomalak:2020zfh}.

The first non-trivial effect from the nuclear Coulomb field appears at three-loop order as shown in \cref{diagram}. In what follows we will consider the point-like limit, $F_{\rm ch}(Q^2) \rightarrow 1$, for which analytic results are known \cite{Wichmann:1956zz}
\begin{equation}
     \mathcal{M}_1^{(1)} =  \frac{Z}{Q_w}\mathcal{M}_0 \sum c_V^{(f)} \Pi_{\rm WK}(Q^2,m_f)~, 
\end{equation}
where $Q_w$ is the weak nuclear charge, $c_V^{(f)}$ is a short-hand for the vectorial Wilson coefficients that appear in Eqs.\ (2.4) and (2.5) of \cite{Tomalak:2020zfh}, and the subscript ``WK'' stands for Wichmann-Kroll \cite{Wichmann:1956zz}. The Wichmann-Kroll contribution is typically studied in coordinate space and can be expressed as a weighted integral over Yukawa-like contributions  \cite{Blomqvist:1972ddn}. Fourier transforming to momentum space, a similar representation may be found for the vacuum polarization
\begin{equation} 
    \label{Pi-WK}
    \begin{split}
    \Pi_{\rm WK}(Q^2,m_f) = \frac12\int_0^\infty & \dd \zeta \frac{Q^2}{Q^2+4\zeta^2m_f^2} \frac{1}{\zeta^4}\\
    & \hspace{-16pt}\qty[ F(\zeta) - \frac{\pi^2}{12} \sqrt{\zeta^2-1}\Theta(\zeta-1)] ~.
    \end{split}
\end{equation}
The function $F(\zeta)$ has relatively lengthy integral representation which can be found in the literature (see e.g.\ pages 99-100 of \cite{Blomqvist:1972ddn}).\footnote{It is useful to note that $F(\zeta) \sim O(\zeta^7)$ for small $\zeta$ such that \cref{Pi-WK} is a well defined convergent integral.}

In the limit of a massless fermion, or equivalently when $Q^2 \gg m_\ell^2$, results are known to all orders in $Z\alpha$ \cite{Brown:1974nsa,Brown:1974fg}.  Structure dependent corrections have also been investigated in for $QR \lesssim 1$ in this limit \cite{Brown:1975rj}. In what follows we apply these well known results to CEvNS and quantify the size of higher order Coulomb corrections for this process from closed lepton loops.

\section{Results in the point-like limit \label{Results} } 
We have numerically computed the Wichmann-Kroll vacuum polarization using the integral representation given in \cref{Pi-WK}. Corrections to CEvNS turn out to be small. For the sake of simplicity, we quantify this with the lepton-flavor asymmetry defined in \cite{Tomalak:2020zfh}. Defining the total vacuum polarization via
\begin{equation}
    \Pi= \Pi_{1} + (Z\alpha)^2  \Pi_{\rm WK } + \ldots, 
\end{equation}
with $\Pi_1$ the leading-in-$\alpha$ contribution, 
the flavor asymmetry in CEvNS is then given by 
\begin{equation}
    \begin{split}
    \mathcal{A}_{\ell,\ell'}&= \frac{\dd\sigma_{\nu_\ell} -  \dd\sigma_{\nu_{\ell'}}}{\dd\sigma_{\nu_\ell}}\\
   &=
    \frac{4\alpha }{\pi} \frac{Z}{Q_w} \qty[\Pi(Q^2,m_\ell)-\Pi(Q^2,m_{\ell'}) ]~.
    \end{split}
\end{equation}
This observable has the advantage of being scheme independent with no residual scale dependence.  The Wichmann-Kroll contribution to $\Pi(Q^2,m_\ell)$ shifts this ratio at $O(Z^2\alpha^2)$. We express this dependence as 
\begin{equation}
    \mathcal{A}_{\ell,\ell'} = 
    \frac{4\alpha }{\pi} \frac{Z}{Q_w} \qty(\mathcal{A}_{\ell,\ell'}^{(0)} + (Z\alpha)^2     \mathcal{A}_{\ell,\ell'}^{(1)}+ \ldots)
\end{equation}
The numerical size of these effects is shown in \cref{asym-shift-table} for three different values of $Q^2$. One can clearly see that coefficient of three-loop $O(Z^2\alpha^3)$ correction is much smaller than that of the vacuum polarization. Similar conclusions apply to the total cross section and indeed any other observable.

\begin{table}[t]
\caption{%
The CEvNS flavor asymmetry for $Q^2=(1~{\rm MeV})^2$, $(10~{\rm MeV})^2$, and $Q^2=(30~{\rm MeV})^2$ in the point-like limit. The result is essentially $Q$-independent above a few MeV. In all cases the Coulomb corrections are very small (even when $Z\alpha$ is not). The $\mu-\tau$ asymmetries are completely negligible. 
\label{asym-shift-table}
}
\begin{ruledtabular}
\begin{tabular}{clccc}
$Q$ &\textrm{Asymmetry }&
{$\mathcal{A}_{\ell,\ell'}^{(0)}$}&
\multicolumn{1}{c}{$\mathcal{A}_{\ell,\ell'}^{(1)}$}\\
\colrule
    ~&  $\ell=e$ ~~$\ell'=\mu$ &  $3.36$ &  $0.02$ \\
    $1~{\rm MeV}~$ ~&$\ell=e$ ~~$\ell'=\tau$  &  $5.24$ &  $0.02$ \\
    ~&   $\ell=\mu$ ~~$\ell'=\tau$  &  $1.88$ &  --- \\
    \colrule
     ~&  $\ell=e$~~$\ell'=\mu$  &  $2.12$ &  $0.03$ \\
    $10~{\rm MeV}~$~&$\ell=e$ ~~$\ell'=\tau$  &  $4.00$ &  $0.03$ \\
    ~&   $\ell=\mu$ ~~$\ell'=\tau$  &  $1.88$ &  --- \\
    \colrule
        ~&  $\ell=e$ ~~$\ell'=\mu$ &  $1.40$ &  $0.03$ \\
    $30~{\rm MeV}~$ ~& $\ell=e$ ~~$\ell'=\tau$  &  $3.28$ &  $0.03$ \\
    ~&   $\ell=\mu$ ~~$\ell'=\tau$  &  $1.88$ &  --- \\
\end{tabular} 
\label{tab:stops}
\end{ruledtabular}
\end{table}

\begin{figure} 
    \vspace{12pt}
    \includegraphics[width=1\linewidth]{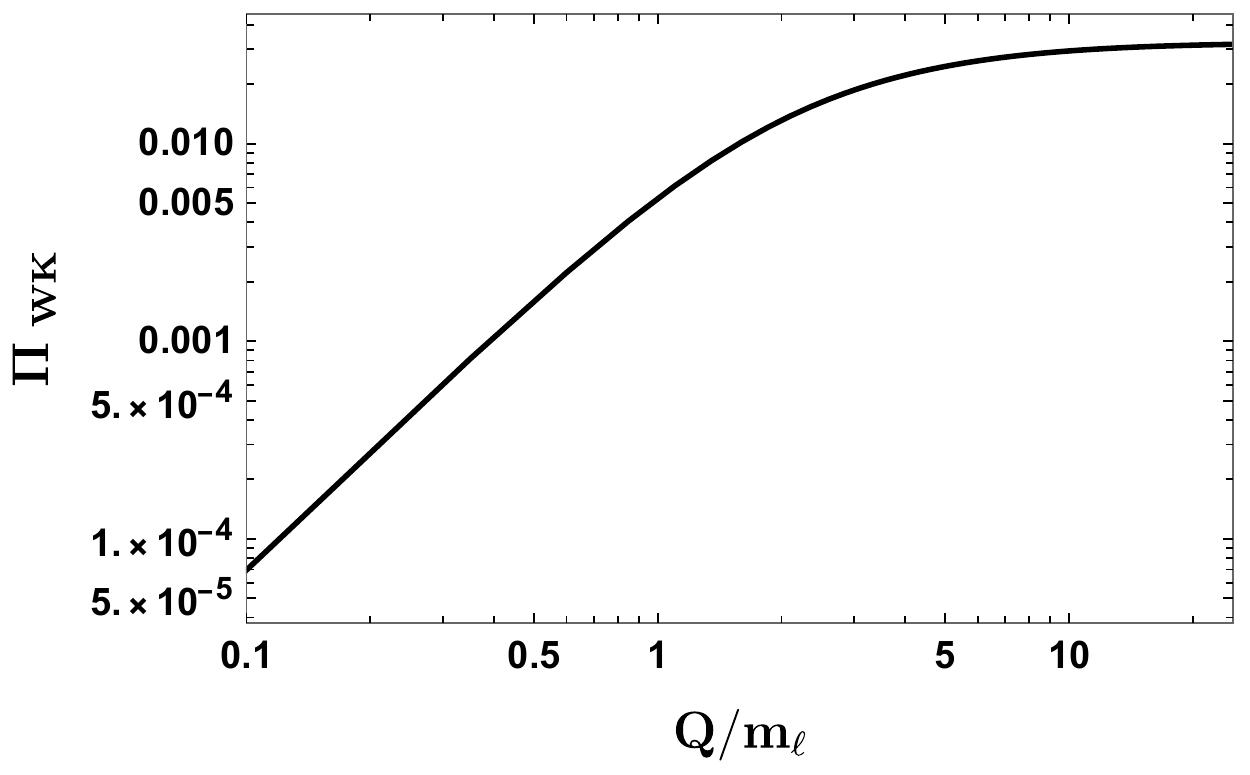} 
    \caption{Numerical calculation of the Wichmann-Kroll vacuum polarization as a function of $Q/m_\ell$. One clearly 
    sees the Coulomb corrections are always much smaller than naive power-counting in $Z\alpha$ 
    would suggest. \label{WK_graph} } 
\end{figure}

Let us now discuss why the corrections are small for heavy leptons, ($\mu$ and $\tau$) v.s.\ light dynamical ones (the electron). All contributions involving 3 or more photon attachments to the lepton loop are UV-convergent,\footnote{This is not obvious by superficial power counting but is guaranteed by gauge invariance \cite{Karplus:1950zz,Brown:1974fg,Liang:2011sj}.} and must decouple as $1/m_\ell^2$ in the limit that $m_\ell \rightarrow \infty$. This is in contrast to the vacuum polarization where a residual logarithmic dependence, correcting for the running value of $\alpha$, persists. This then guarantees that corrections are parametrically suppressed by $(Z\alpha)^2 Q^2/m_\ell^2$. For electrons, $Q^2 \gg m_e^2$ and so this argument does not apply, however the short-distance coefficient of the Wichmann-Kroll potential is small being given by 
\begin{equation}
    \label{WK_lim}
    \lim_{Q\rightarrow \infty} \Pi_{\rm WK}(Q/m_\ell)= \frac{1}{2} \qty[\frac{\pi ^2}{6}-\frac{2}{3}  \zeta (3)-\frac{7}{9}]\approx 0.03 ~. 
\end{equation}
This suppression has been shown to persist to all orders in $Z\alpha$ for the hierarchy $m_e \ll Q \ll 1/R$ where $R$ is the nuclear radius; we discuss finite size effects in the next section. A cursory examination of \cref{Pi-WK} and the observation that $0\leq Q^2/(Q^2 + 4 m_\ell^2\zeta^2)\leq 1$ guarantees that $\Pi_{\rm WK}(Q^2)$ is a monotonically increasing function of $Q^2$; this is shown explicitly in \cref{WK_graph}. Therefore, since corrections are small in the limit that $Q^2\gg m_e^2$ they are small for all values of $Q^2$. 

\section{Additional corrections \label{Discussion} } 
In the previous section we have presented results for higher order corrections to the vacuum polarization in the point-like limit. We focused on leptonic contributions, however in the full CEvNS cross section hadronic loops also contribute. One may ask how hadronic contributions to the vacuum polarization and finite size corrections may alter our results. In this section we will argue these effects are never large, nor are finite size corrections from the charge distribution of the nucleus. 

Hadronic contributions will always be small on dimensional grounds.  For the same reason that heavy leptons decouple, hadrons must also. Since the lightest hadron is $m_\pi$ the effect of hadronic loops will always be suppressed by the same factor of $Q^2/m_\pi^2$. This ratio is typically smaller than $0.1$ for CEvNS kinematics. It is possible the Coulomb corrections from pion loops may be larger than the leptonic contribution, however they will never be ``large'' in so far as $(Z\alpha)^2 Q^2/m_\pi^2 \ll 1$. 

Our discussion so far has focused on the point-like limit. Finite size effects have been previously studied in the context of mesonic and muonic atoms where a coordinate space representation for the potential was derived \cite{Brown:1975rj}, valid to all orders in $\alpha (Z\alpha)^{2n+1}$, for $r \gg R$; finite size effects were found to be small as one would expect. For neutrino energies above $30~{\rm MeV}$ or so, a sizeable fraction of the CEvNS cross section comes from regions where $QR\sim 1$ due to the phase-space enhancement of large-$Q$ regions. In these regions of phase space the point-like calculations presented above do not strictly speaking apply, but should provide a reasonable estimate of the radiative corrections. Since they are so small we do not attempt to compute the three-loop integral or to construct modified Greens functions for an extended nuclear charge distribution. In addition to anticipating that the corrections will never be larger than in the point-like limit, we note that in the regime where $QR \sim 1$ uncertainties due to the tree-level nuclear form factor dominate the error budget contributing an uncertainty at the level of a few percent (see Table 2 of \cite{Tomalak:2020zfh}).

Finally, one may wonder if the cancellation in \cref{WK_lim} is accidental such that higher order terms may be important. The analysis of \cite{Brown:1974fg} shows this not to be the case; all higher order corrections are similarly suppressed by a factor of roughly $1/100$.

\section{Conclusions \& Outlook \label{Conclusions} } 
In this work the leading Coulomb corrections to CEvNS, directly related to the Wichmann-Kroll potential, have been calculated in the limit of a point-like nucleus. The corrections are found to be small. The influence of finite size corrections, mixed hadronic correlators, and other SM effects not explicitly computed have been estimated and will not affect the CEvNS cross section at a level relevant for experiments (i.e.\ they are sub-permille). 

In a broader context, this work demonstrates that Coulomb corrections are under control, a necessary requirement for a precision CEvNS program. This compliments other recent additions to the literature such as an estimate of inelastic invisible contributions to the cross section \cite{Dutta:2022tav}, and percent-level determinations of neutrino fluxes \cite{COHERENT-FLUX,Tomalak:2021lif}. Further work on hadronic contributions, via the mixed correlator $\Pi_{3\gamma}$ discussed in \cite{Tomalak:2020zfh}, and a reduction of uncertainties in the neutron and proton distributions inside nuclei offer a path towards a sub percent prediction of the CEvNS cross section. 

\section*{Acknowledgements} 
I would like to thank Oleksandr Tomalak, Richard Hill, and especially Robert Szafron for helpful discussions, and both Robert Szafron and Oleksandr Tomalak for reading this manuscript before completion. This work was instigated at the Mainz Institute for Theoretical Physics, and I thank them for their hospitality. This work is funded by the Neutrino Theory Network Program Grant under Award Number DEAC02-07CHI11359 and the US DOE under Award Number DE-SC0020250.
This work is supported in part by the U.S. Department of Energy, Office of Science, Office of High Energy Physics, under Award Number DE-SC0011632 and by the Walter Burke Institute for Theoretical Physics. 
\bibliography{biblio.bib}
\end{document}